# FUSE: A Partitioned Field-Exchange Framework for Coupling Physics Simulations in FEBio

Steve A. Maas[+1], Farhan Muhib[+1], Jeffrey A. Weiss[1*]

[1]Department of Biomedical Engineering, and

Scientific Computing and Imaging Institute,

University of Utah, Salt Lake City, UT

+ co-first authors

*Corresponding Author:
Jeffrey A. Weiss
Department of Bioengineering
University of Utah
72 South Central Campus Drive, Rm. 2646
Salt Lake City, UT 84112
jeff.weiss@utah.edu

**Abstract**

Computational biomechanics increasingly requires models that combine mechanics, transport, chemistry, and biological regulation across different spatial and temporal scales. The FEBio simulation software (Finite Elements for Biomechanics and Biophysics) provides extensive open-source capabilities for modeling these processes using monolithic approaches. However, assembling independently developed physics models into reproducible coupled workflows remains challenging. Existing approaches often require custom scripts or external software pipelines, which can limit model reuse and complicate development. We present FUSE, the FEBio Unified Simulation and Exchange framework, a partitioned coupling plugin that enables separately defined FEBio models to communicate through structured field exchange. FUSE is designed for problems that are best solved independently, particularly when fast mechanical responses influence slower biological or chemical evolution. The framework uses a time-decoupled strategy in which a primary model advances on the longer time scale, while one or more secondary models are repeatedly initialized, supplied with updated fields, solved over shorter time horizons, with results returned to the primary model. Field exchange utilizes existing FEBio data maps, output fields, and user-specified filters, allowing coupled workflows to be constructed without modifying the underlying solvers. The framework was able to reproduce reference coupled solutions while handling bidirectional transfer, spatial field mapping, and filtered exchange of model variables. Example applications demonstrated coupling between mechanical loading and chemical degradation in injured cartilage and interaction between biological tissue formation and mechanical feedback during bone healing. By separating coupling logic from physics implementation, FUSE provides a practical mechanism for building maintainable multiphysics workflows within FEBio.

## 1. Introduction

Many problems in computational biomechanics are inherently multiphysics and multiscale, and accurate prediction increasingly depends on coordinated treatment of mechanics, transport, chemistry, and cell-mediated regulation rather than isolated treatment of a single process [1-3]. Importantly, the physical processes within a single multiphysics model often unfold over dramatically different time scales. One example arises in biological tissues, where catabolic and anabolic processes are regulated through the combined interaction among mechanical, chemical, and biological systems [1, 2, 4]. In hydrated tissues, deformation alters fluid flow, solute transport, and reaction kinetics, while evolving chemical and cellular states regulate growth, remodeling, damage, and degeneration [5-7]. This is particularly evident in load-bearing tissues during growth and healing, where transport and chemistry do not simply accompany deformation but participate directly in setting tissue state and function [5, 7]. Furthermore, mechanical loading and the associated strain fields typically evolve over relatively short time scales, whereas downstream biological processes such as cell migration, proliferation, apoptosis, and matrix turnover may evolve over hours, days, or longer in applications such as fracture healing, angiogenesis, and tissue degeneration [7-9].

This scientific need is accompanied by a practical computational one. The physics models needed to represent various aspects of a biological system are often developed independently, validated in distinct settings, and implemented using different numerical formulations and software platforms [4, 10]. Even when each subsystem model is mature, linking them can require custom solutions for field transfer, time synchronization, data management, and iterative coupling [10-12]. The burden grows further when coupled subsystems rely on different spatial discretizations or distinct time-stepping requirements, which is common when slow biological evolution is linked to rapidly occurring mechanics [10, 12]. The main obstacle in such settings is therefore often not the absence of a governing model for a given process, but the lack of a reusable framework that allows separately developed physics models to interact within a common workflow [3, 11].

Within the field of biomechanics simulation using the finite element method, FEBio was developed to address part of this need by providing an open-source finite element environment tailored to the mechanical simulation of biological tissues and biophysical systems [13]. It has become a highly popular research platform for computational biomechanics, with over 100,000 downloads of FEBio installation packages, over 29,000 registered users, and cited in more than 1,000 peer-reviewed publications [14]. FEBio has evolved into a comprehensive multiphysics platform supporting coupled analyses of mechanics, transport, chemistry, growth and remodeling, computational fluid dynamics, and fluid-structure interaction within a common software environment [6, 15-18]. The FEBio plugin framework further increased extensibility by allowing new functionality to be distributed as dynamically linked libraries [19]. It has enabled specialized applications ranging from reaction-diffusion modeling to collagen-guided angiogenesis and multiscale systems biology simulations [8, 19-21]. Despite FEBio's versatility, there remains an important need for physics models to have a systematic way to exchange data and advance in time together.

In multiphysics computation, coupled systems are commonly treated either monolithically, with all governing equations assembled and solved in a single computational environment, or in a partitioned form, with subsystems solved separately and synchronized through data exchange [10, 12]. Monolithic formulations, as implemented in FEBio's multiphasic framework [6, 15], can be advantageous when coupling is strong, but they require all physics to be implemented with the same spatial and temporal discretizations and the same nonlinear solver framework. In many situations, this can be disadvantageous. For instance, in mechanobiology simulations, mechanical loading occurs over seconds to minutes, while chemical signaling, cell migration, differentiation, and tissue remodeling/growth occur over hours to days. In these cases, a monolithic approach often results in poor conditioning of the stiffness matrix, excessive timestep requirements, large nonlinear systems, and expensive linear solves, and it complicates the exploitation of the multi-rate structure of distinct physical processes. The large coupled systems arising from implicit time integration often require direct linear solvers for robustness, with memory and computational cost increasing rapidly due to fill-in during factorization [22]. The global timestep is governed by the most demanding physical process, leading to inefficient integration when coupled phenomena evolve on different time scales.

In contrast, partitioned approaches are appropriate for weakly coupled physics and allow each subsystem to retain its own formulation and numerical method, improving flexibility and model reuse [10, 12]. Partitioned strategies decompose the problem into physics subsystems, thereby reducing the size and complexity of individual linear systems. Partitioned approaches allow different subsystems to advance using time steps appropriate to their own dynamics [23]. Broadly coupled ecosystems such as preCICE, together with programmable finite element platforms such as FEniCS, demonstrate the value of reusable infrastructure, but their general-purpose nature requires additional configuration for coupling and solver-side integration [11, 24, 25]. For multiple FEBio-centered simulation workflows in which physics models already share a common mesh, data structures, and field definitions, the remaining need for coupling is therefore for a systematic, reusable, easy-to-implement way to exchange fields between separately defined physics models.

To address this gap, we developed a novel FEBio plugin called FUSE (FEBio Unified Simulation and Exchange), a generalized coupling framework designed to meet this need within the FEBio ecosystem. The partitioned architecture allows coupled analyses to remain separately formulated while participating in a coordinated simulation. The objective of the present work is to describe the formulation and software design of FUSE and to demonstrate how the framework can support reproducible two-way data exchange among models that differ in physics, constitutive assumptions, or temporal resolution. Given the broad adoption of FEBio within the biomechanics community, this framework will provide existing users with a practical mechanism to extend their models to coupled and multiscale applications without requiring substantial redevelopment or migration to external coupling infrastructures. By formalizing these interactions within a common workflow, FUSE will reduce case-specific implementation effort and expand the range of coupled biomechanical simulations that can be assembled from existing capabilities of FEBio. This is particularly useful for problems in which each subsystem is best solved separately. For example, the cartilage degradation framework of Eskelinen et al. required COMSOL for biochemical degradation, ABAQUS for biomechanical loading, and MATLAB to exchange degradation fields between the two models [26]. FUSE provides an FEBio-centered alternative to this type of multi-software workflow. The mechanical and chemical models can remain separate, but a single reusable coupling plugin handles field exchange, filtering, and repeated execution.

FUSE has practical value for the broader FEBio user community by lowering the implementation barrier to coupled mechanobiological modeling. Many FEBio users have already built reliable single-physics or internally coupled models using existing FEBio modules, materials, reactions, and boundary conditions [27-29]. However, extending these models to communicate with separate analyses often requires custom scripts, external data processing, or problem-specific code. FUSE formalizes this exchange through FEBio data maps, output variables, and user-defined filters, allowing users to construct coupled workflows while continuing to work within standard FEBio model files and the FEBio plugin ecosystem [13, 15, 19]. Further, the physics models in FEBio are open-source and easily extensible. In this way, FUSE can help users prototype multiscale and multiphysics models more quickly, reduce dependence on case-specific coupling scripts and closed-source commercial software, and improve reproducibility by keeping the fully coupled workflow in an explicit FEBio-compatible configuration.

As with all software that is part of the FEBio project, FUSE is open source and available in a GitHub repository [30].

## 2. Methods

### 2.1. Time-decoupled Strategy

In this paper, we focus mostly on problems in which one subsystem evolves on a relatively slow time scale (the primary system), while one or more additional subsystems evolve on much shorter time scales (the secondary systems). FUSE solves these problems using a *time-decoupled* strategy that can be viewed as a nested staggered (or sequential) partitioned solution scheme. The approach is motivated by the observation that the state of the primary system changes little during the evolution of the secondary systems, allowing the primary state to be treated as constant while each secondary model is solved. The present implementation corresponds to an explicit partitioned coupling scheme, since information is exchanged only once during each primary time step and no iterations are performed to enforce convergence between the coupled models. Unlike conventional explicit partitioned schemes, in which all coupled models advance together through physical time, the secondary models in FUSE are reinitialized and solved independently over their complete time horizon during each primary time step. This execution pattern naturally accommodates applications in which short-term physical processes repeatedly determine the long-term evolution of a slower biological system.

The implementation of the time-decoupled strategy in FUSE does not require data exchange to occur at every primary model time step. Instead, users may specify an *exchange interval*, which determines how many primary time steps elapse between successive field exchanges and secondary model solves. This allows the primary model to advance through multiple time steps before the secondary models are re-evaluated, potentially reducing computational cost when the primary state changes slowly relative to the coupling interval.

The stability and convergence of the time-decoupled strategy depend on the physical processes being coupled and on the degree of separation between their characteristic time scales. Consequently, a general theoretical analysis is difficult to obtain. For general coupled problems, explicit sequential partitioned algorithms are typically first-order accurate in time, with stability depending on the strength of the coupling and the size of the primary time step. As the coupling becomes stronger or the characteristic time scales become less well separated, the numerical solution may become increasingly sensitive to the primary time step or even diverge. Furthermore, choosing a larger exchange interval reduces computational cost but may also reduce accuracy by delaying propagation of information between the coupled models. Nevertheless, for problems exhibiting sufficient time-scale separation, this explicit time-decoupled scheme can be successfully applied. The verification and example problems presented below demonstrate that the proposed strategy accurately reproduces reference solutions for representative biomechanics applications.

### 2.2. The FUSE plugin

FUSE was developed using the FEBio plugin framework. FEBio supports many types of plugins, ranging from materials, boundary conditions, and loads to complete physics solvers [31]. FUSE is implemented as a *task* plugin, which differs from other plugin types in that it controls the overall execution flow of FEBio [32]. In other words, the task plugin defines how FEBio and its various solvers are orchestrated during a simulation. This gives developers the flexibility to implement complex workflows, such as inverse modeling algorithms or procedures that require repeated forward simulations. Since FUSE repeatedly advances coupled forward models operating on different time scales, the *task* plugin architecture was a natural fit for its implementation.

The FUSE task plugin manages multiple FEBio models consisting of one primary model and one or more secondary models. Its main application is to solve models with decoupled time scales: the primary model corresponds to the system with the largest time scale, while the secondary models operate on smaller time scales. The coupled system is solved using the *time-decoupled* strategy, as explained above. The primary model is advanced as a standard forward simulation. At the beginning of each primary time step, all secondary models are solved first. Before solving the secondary models, they are reset to their initial configuration, and any field data needed from the primary model is transferred to them. The secondary models are then solved until completion. Once the secondary models have been solved, any necessary field data is transferred back to the primary model, after which the primary model time step is advanced. This process is repeated until the primary simulation is complete (Fig. **1**).

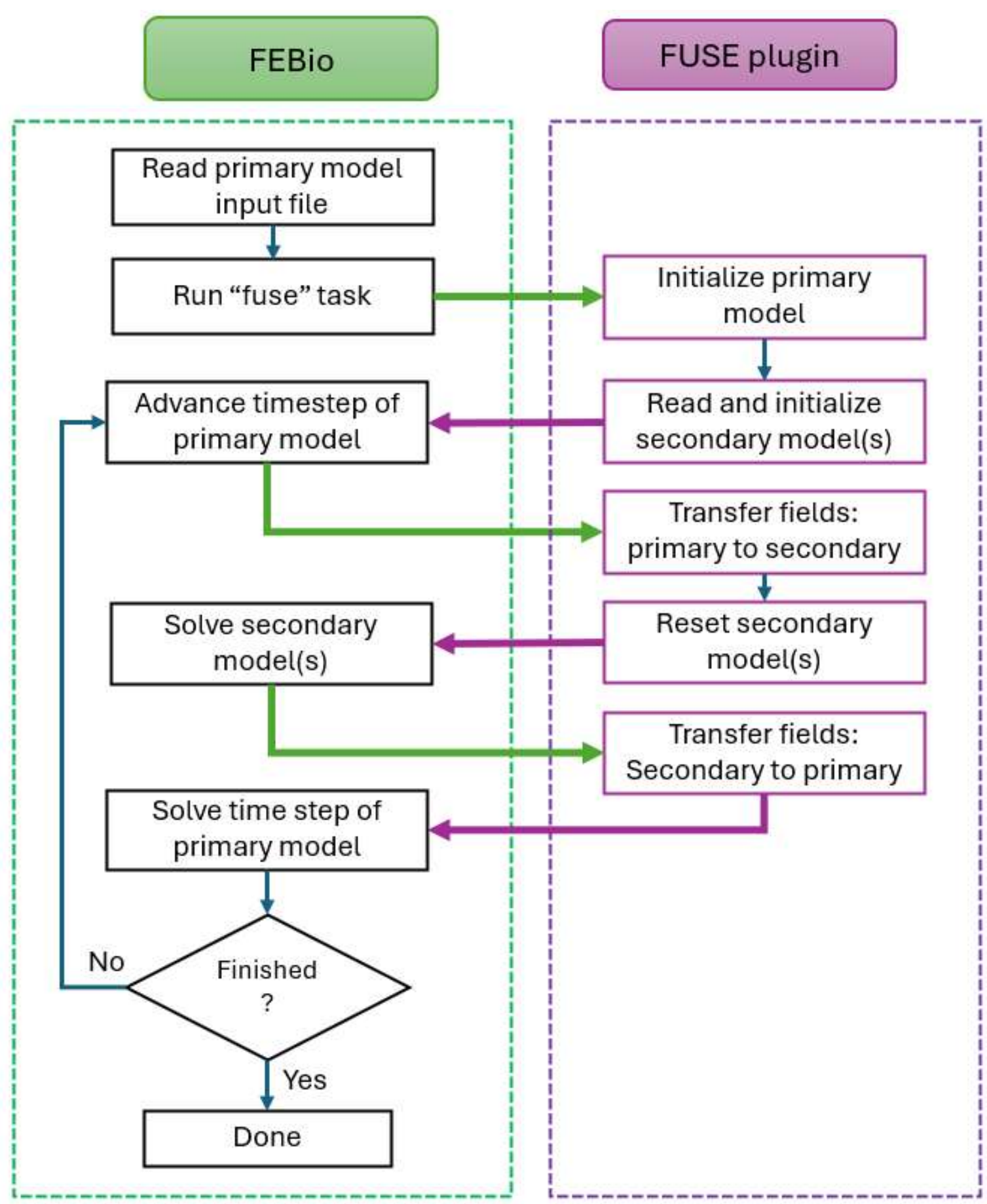

**Figure 1**: Execution workflow and the coordination of the FUSE task plugin within FEBio. FUSE initializes the primary and secondary FEBio models and then coordinates with FEBio as it advances the simulation by iterating over the primary model time steps. At each primary time step, the required fields are transferred to the secondary model(s), which are reset and solved independently to completion. The resulting fields are then transferred back to the primary model before advancing to the next primary time step. This partitioned strategy allows each secondary model to use its own temporal resolution while coupling to the evolving primary simulation.

Data exchange between the primary and secondary models is implemented using the FEBio *data map* feature. Data maps allow users to define spatially varying model parameters, such as position-dependent material densities or surface pressures. Using data maps as the exchange mechanism reuses existing FEBio infrastructure and avoids the need to implement a dedicated exchange framework within the plugin. This approach does require users to prepare FEBio models for use with FUSE by defining data maps for the parameters that will participate in the coupling.

For existing models, this generally requires only minor modifications, as demonstrated in the examples below. Within FUSE, a data map serves as the destination for data received during an exchange operation, while the corresponding data source is defined using any of the standard FEBio output fields. Examples of output fields include the primary solution variables (e.g., displacement or concentration) and internal state variables (e.g., strain, stress or concentration flux). Currently, FUSE assumes that the meshes of the primary and secondary models are identical; that is, they must contain the same nodes and elements in the same ordering. This assumption simplifies communication between models by enabling direct node-to-node or element-to-element mapping, depending on the type of exchanged data.

As part of the data exchange definition, users may specify filters that transform the source data before it is stored in the destination data map. These filters can serve a variety of purposes, such as converting between different unit systems used by the primary and secondary models. Filters are implemented using the one-dimensional function definitions of FEBio, which support several representations, including interpolated point data, mathematical expressions, and predefined functions such as scaling or linear ramps.

The primary and secondary models are defined as standard FEBio input files, and the FUSE task is configured using an XML-formatted *options* file, typically assigned the extension ".opt". This options file defines the secondary models (the primary model is specified on the FEBio command line), the data exchanges, and any configuration parameters for the time-decoupled strategy. See the Supplementary Material document for additional details on configuring and using the FUSE plugin with FEBio.

Since the primary model is solved as a forward model, FEBio will output the standard log and plot files for this model. In addition, FUSE will create a plot file for each secondary model. The plot file is named after the secondary model and contains the final states from the forward solves of that model.

## 3. Verification and Example Problems

The input files for all problems in this section can be found in the FEBio Model Repository, accessible directly through the FEBio Studio UI or via the persistent URL: https://repo.febio.org/permalink/project/148. Although FUSE can accommodate coupling between any FEBio models, the verification and example problems below utilize FUSE to couple a reaction-diffusion model with a mechanical model. The reaction-diffusion model uses the FEBioChem plugin [19, 33]. This coupling is especially relevant to problems in biomechanics that couple mechanics to chemical and growth factor signaling, tissue engineering, biosynthesis binding and degradation, and growth and remodeling.

### 3.1 0D Verification

The 0D problem provides the simplest verification of two-way coupling by removing spatial variation and isolating the FUSE exchange logic. It represents a lumped model of mechanobiological bone adaptation where the bone density, $\rho$, evolves according to a kinetic equation. A scalar strain measure, $\varepsilon$, represents the mechanical response:

$$\varepsilon = \frac{F}{E(\rho)}, \tag{1}$$

where $F$ is the applied force, and $E$ is a density-dependent Young's modulus. The chemical model evolves the concentration-like variable $\rho$ according to:

$$\frac{d\rho}{dt} = k(\varepsilon). \tag{2}$$

Using the following linear assumptions for dependence of $E$ on $\rho$ and $k$ on $\varepsilon$:

$$E(\rho) = E_0 + \alpha\rho , \tag{3}$$

$$k(\varepsilon) = k_0 + k_1\varepsilon, \tag{4}$$

the coupled problem reduces to the single reference ODE that evolves the density in time:

$$\frac{d\rho}{dt} = k_0 + k_1 \frac{F}{E_0 + \alpha\rho} \tag{5}$$

This problem includes a 2-way closed feedback loop, where the chemical state controls the mechanical modulus $E$ and the mechanical strain controls the chemical production rate for $\rho$. This feedback is implemented by assigning the chemical model as the primary model and the mechanical model as the secondary model.

FUSE transfers density $\rho$ from the chemical model to the mechanical modulus map $E$ using:

$$E = 1000 + 500\rho, \tag{6}$$

and transfers the mechanical strain measure $\varepsilon$ back to the rate map $k$ using:

$$k = 0.01 + 100\varepsilon. \tag{7}$$

Both exchanges are performed using the time-decoupled strategy with an exchange interval of one.

We obtained a numerical solution to the ODE equation (5) using MATLAB to serve as a gold standard. The maximum absolute concentration difference between the FUSE solution and the numerical ODE solution was minimal, with an RMS difference of approximately 0.36% of the ODE reference range (Fig. 2a). The agreement between the FUSE/FEBio solution and the independent ODE solution verifies that the FUSE correctly handles bidirectional scalar transfer, linear filters, model synchronization, and update order.

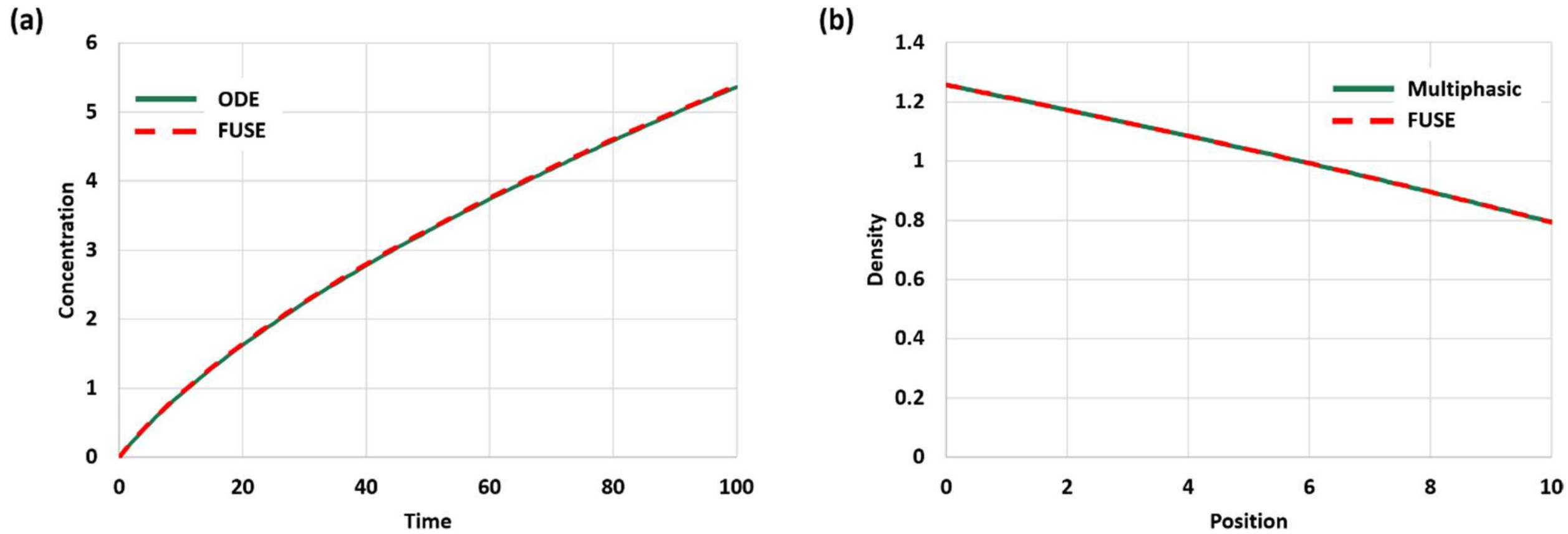

**Figure 2**: Verification of FUSE coupling for reduced-order and spatially resolved mechano-chemical problem. (a) 0D verification problem showing concentration evolution over time, where the FUSE solution agrees closely with the independent ODE reference solution, verifying bidirectional exchange between the chemical state and mechanical response. (b) 1D verification showing the steady-state density distribution along the model domain, where the FUSE solution reproduces the monolithic multiphasic reference solution, verifying spatial data transfer, field mapping, and nonlinear coupling through the FUSE control file. Solid green lines denote reference solutions, and dashed red lines denote FUSE results.

### 3.2 1D Verification

The 1D problem extends the verification from scalar feedback to spatially varying, element-wise coupling. The reference problem is based on the monolithic multiphasic formulation described by Ateshian et al. (2014), in which the material density of a solid-bound molecule (SBM) evolves according to a mechanosensitive production law, and the mechanical stiffness depends on the SBM density [6]. The simplified Huiskes-type reaction rate, $k$, is written as:

$$k = B\left(\frac{\Psi_r}{\rho_r^s} - \psi_0\right), \tag{8}$$

where $\Psi_r$ is the strain-energy density of the solid, $\rho_r^s$ is the SBM density, $\psi_0$ is the specific strain energy at homeostasis, and $B$ is a material constant [34]. The solid material is represented using a Carter-Hayes constitutive model, which is essentially a neo-Hookean material where the Young's modulus $E$ depends on the SBM density via a power law:

$$E(\rho_r^s) = E_0\left(\frac{\rho_r^s}{\rho_0}\right)^{\gamma}. \tag{9}$$

Here, $E_0$ is the initial elastic modulus, $\rho_0$ is the initial density, and $\gamma$ is a material constant. The Huiskes remodeling framework and Carter-Hayes density-dependent stiffness law are established components of the FEBio multiphasic/solid material descriptions [34-37]. This framework has been used to simulate bone adaptation to mechanical loading.

The main numerical problem is that the applied load produces a nonuniform strain-energy density along the prismatic bar, which, in turn, leads to a nonuniform reaction rate and a spatially varying modulus. In the FUSE version, the chemical and mechanical problems are solved as separate FEBio models on the same geometry and mesh. The chemical concentration, $c$, representing the density of the SBM, is mapped to the mechanical modulus, $E$, through the filter specified in the *options* file:

$$E = 10000c^2, \tag{10}$$

which corresponds to the power-law density-stiffness relation with $E_0 = 10000$, $\rho_0 = 1$, and $\gamma = 2$. Conversely, the mechanical strain-energy density, $\Psi_r$, is transferred directly to the chemical model as the rate-control map.

The FUSE result was compared with the monolithic multiphasic solution by examining the steady-state density profile along the bar. The two profiles agreed closely over the full domain; based on the plotted data provided, the maximum absolute density difference was $1.94\times10^{-3}$, and the normalized RMS difference was approximately 0.10% of the reference density range (Fig. 2b). This agreement verifies that FUSE can reproduce the monolithic coupled multiphasic solution when the coupling terms are represented through field exchanges between separated chemical and mechanical models. Thus, the 1D test verifies not only bidirectional transfer, but also element-wise spatial mapping, nonlinear filtering, and steady-state mechano-chemical feedback.

### 3.3 Example Problem: Cartilage Damage

This example demonstrates the use of FUSE to couple two FEBio analysis models operating on different characteristic time scales — a biphasic mechanical loading model and a reaction-diffusion chemical degradation model. The example was motivated by previous studies of mechanically injured articular cartilage, which identified lesion-induced strain concentrations and the diffusion of inflammatory cytokines as mechanisms contributing to proteoglycan (PG) and fixed charge density (FCD) loss [26, 38]. The objective of this test problem was to demonstrate how FUSE can exchange spatially varying state variables between mechanical and chemical analyses during an iterative mechanochemical degradation simulation.

Articular cartilage was represented as a hydrated extracellular matrix composed of a PG-rich ground substance reinforced by a fibrillar collagen network. In this simplified model, PG represented the intact FCD-bearing matrix, while degraded PG (dPG) represented local matrix loss. The geometry consisted of a 3 mm wide and 1 mm thick rectangular cartilage explant with a superficial lesion on the top surface, discretized with TET10 elements (Fig. 3). The lesion had a depth of 0.144 mm and a width of 0.316 mm, based on previously reported cartilage injury geometries [26, 38]. The model was implemented as a thin three-dimensional domain with an out-of-plane thickness of 0.05 mm, and the front and back faces were constrained in the out-of-plane direction to represent plane-strain boundary conditions. A corresponding intact geometry without a lesion was also generated.

The chemical model represented a simplified interleukin-1 (IL-1)-driven PG degradation pathway. Three species were included: intact PG; IL1; and dPG. [PG] was normalized so that [PG] = 1 corresponded to the intact initial matrix, and [dPG] = 0 corresponded to no initial degradation. A constant IL-1 concentration of 1 ng/ml was prescribed on the top, lateral, and lesion surfaces, while no IL-1 flux was allowed through the bottom, front, or back surfaces. IL-1 diffused into the tissue with an effective diffusivity of $4.0 \times 10^{-7}$ mm²/s. PG and dPG were treated as effectively matrix-bound. Chemical degradation was represented using a simplified catalytic reaction in which IL-1 promoted conversion of intact PG into degraded PG without being consumed (Eqn. 11).

$$PG \xrightarrow{IL-1} dPG \tag{11}$$

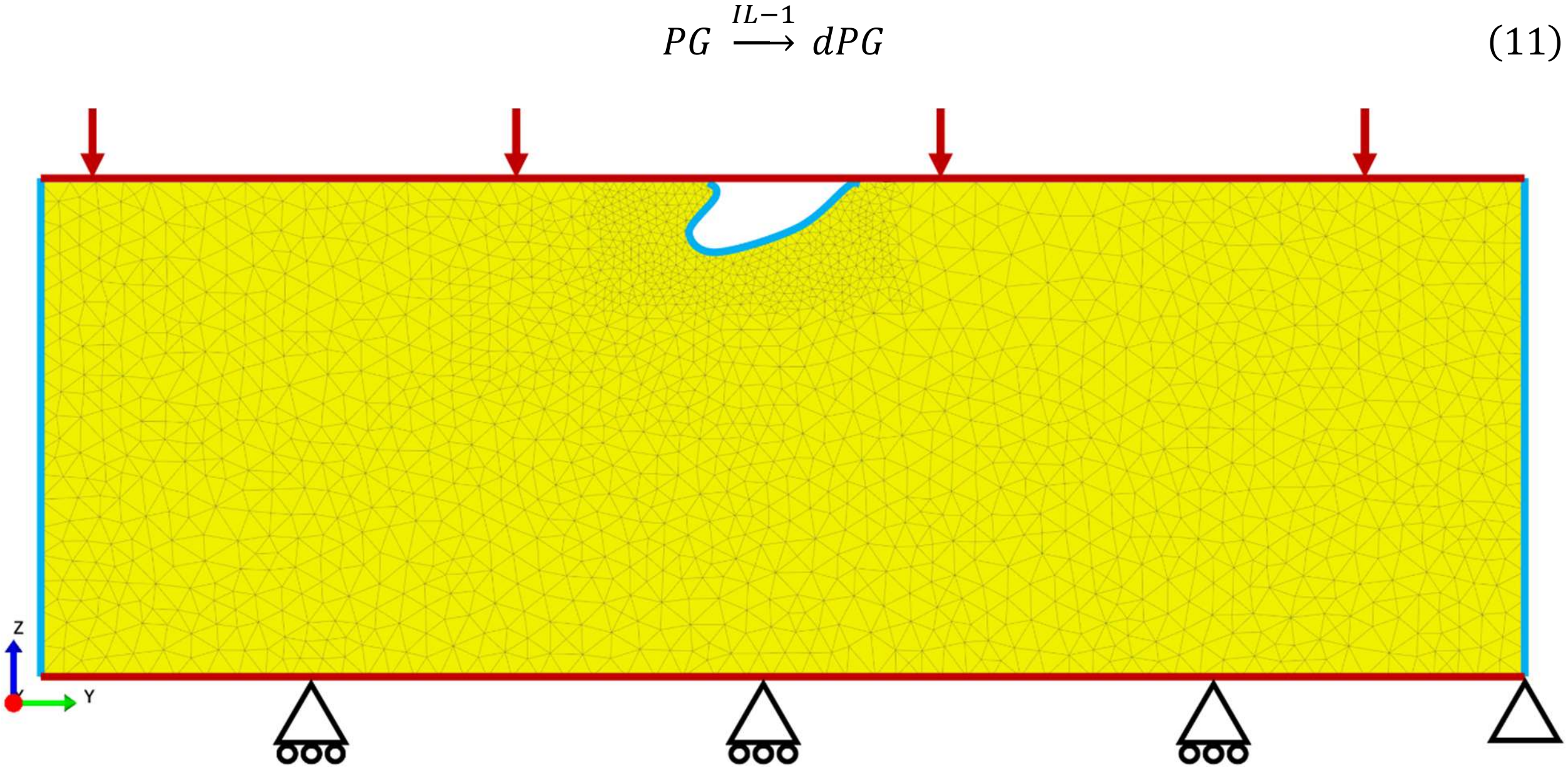

**Figure 3**: Geometry and boundary conditions for cartilage damage model. The cartilage is situated on an impermeable platen on the bottom and strain is applied via a top platen. While the bottom face is fixed vertically, one bottom-left node was constrained in all directions. The blue marked boundaries were assigned to have zero fluid pressure, making them freely permeable to fluid flow.

The mechanical module represented cartilage as a fibril-reinforced biphasic material. The solid volume fraction was 0.225, the permeability was 0.0013 $mm^4/(N\cdot s)$, and the nonfibrillar matrix was assigned an elastic modulus of 0.16 MPa and Poisson's ratio of 0.42 [38]. Collagen reinforcement was represented using two simplified *fiber-exp-linear* fiber families oriented at $0^\circ$ and $-120^\circ$ relative to the local cartilage coordinate system, with each family assigned a fiber stiffness parameter of 10 MPa, giving a total nominal collagen reinforcement contribution of 20 MPa [38, 39]. The bottom surface was fixed vertically, and one bottom node was constrained horizontally to remove rigid-body motion (Fig. 3). A 15% compressive displacement was applied to the top surface using a 1 Hz triangular waveform. Each mechanical loading analysis consisted of two cycles over 2 s, and the maximum shear strain field was extracted at the second compression peak [26]. Finally, zero fluid pressure was assigned on the side faces and the lesion, making them freely permeable to fluid flow (Fig. 3).

FUSE was used to exchange information between the two models. We defined the chemical model as the primary and the mechanical model as the secondary. In each coupling cycle, the mechanical model was run first to compute the maximum shear strain field ($\varepsilon$). This field was mapped to the chemical module and converted into a strain-dependent PG degradation rate ($k_\varepsilon$). The chemical module was then advanced for 5 hours before returning to the mechanical module for the next loading cycle. This staggered workflow allowed short-time-scale mechanical loading and longer-time-scale chemical degradation to be coupled iteratively. Mechanical degradation was activated only when the local maximum shear strain exceeded a prescribed threshold ($\varepsilon_{thres}$) of 0.25, a value selected to mimic simulation results in previous studies [26, 40]. Below this threshold, there was no mechanically driven PG loss:

$$PG \overset{k_\varepsilon}{\rightarrow} dPG \,. \tag{12}$$

Above the threshold, the mechanical degradation factor ($R_{deg}$) was calculated from the local maximum shear strain:

$$R_{deg} = \frac{1}{3}\sqrt{\varepsilon - \varepsilon_{threshold}}\,. \tag{13}$$

The dimensionless degradation factor was converted into an equivalent first-order reaction-rate constant, $k_\varepsilon$, over the mechanochemical coupling interval ($\Delta t = 5\ h$) [26]:

$$k_\varepsilon = -\frac{\ln(1 - R_{deg})}{\Delta t}\,. \tag{14}$$

The remaining PG concentration [PG] was then used to update the effective modulus $E$ of the cartilage ground matrix before the next mechanical analysis. The modulus decreased linearly with PG loss, with a lower bound, 0.02 MPa, imposed to avoid numerical convergence issues:

$$E = 0.02 + 0.14\ \times\ [PG]\,. \tag{15}$$

The coupled models were first run with mechanically driven PG degradation only, with IL-1 diffusion omitted to isolate the effect of mechanical stimulus on PG loss. A second set of models was then run with combined biomechanical and biochemical degradation, in which physiological loading was applied while IL-1 was allowed to diffuse into the cartilage. The intact model showed little mechanically driven PG loss under cyclic compression, indicating that physiological loading alone did not trigger mechanical degradation in the absence of a lesion (Fig. *4*a). In contrast, the injured model produced localized PG loss beneath the lesion, where maximum shear strain exceeded the prescribed threshold. PG loss occurred rapidly during the first few days of repeated loading, resulting in 55% PG loss after 4 days, and then progressed at a slower rate (Fig. *4*b). When IL-1 diffusion was included, PG loss also developed near the exposed free surfaces. In the intact model, degradation was highest (35% at day 4) near the outer surfaces and decreased toward the tissue interior (2% at the center) (Fig. 4c). In the injured model, the lesion amplified the degradation response by introducing both an additional IL-1 diffusion surface and a localized mechanical strain concentration. As a result, the largest PG loss occurred near and beneath the lesion (86% at day 4), where biochemical and biomechanical degradation mechanisms overlapped (Fig. 4d).

Overall, this test problem demonstrates that FUSE can coordinate sequential mechanical and chemical FEBio analyses, transfer spatially varying fields between modules, and simulate coupled mechanochemical tissue degradation over disparate time scales. The example also illustrates a key mechanobiological concept from previous cartilage degradation studies, where the presence of a lesion can localize mechanically driven PG loss, whereas inflammatory cytokine diffusion produces broader surface-associated degradation.

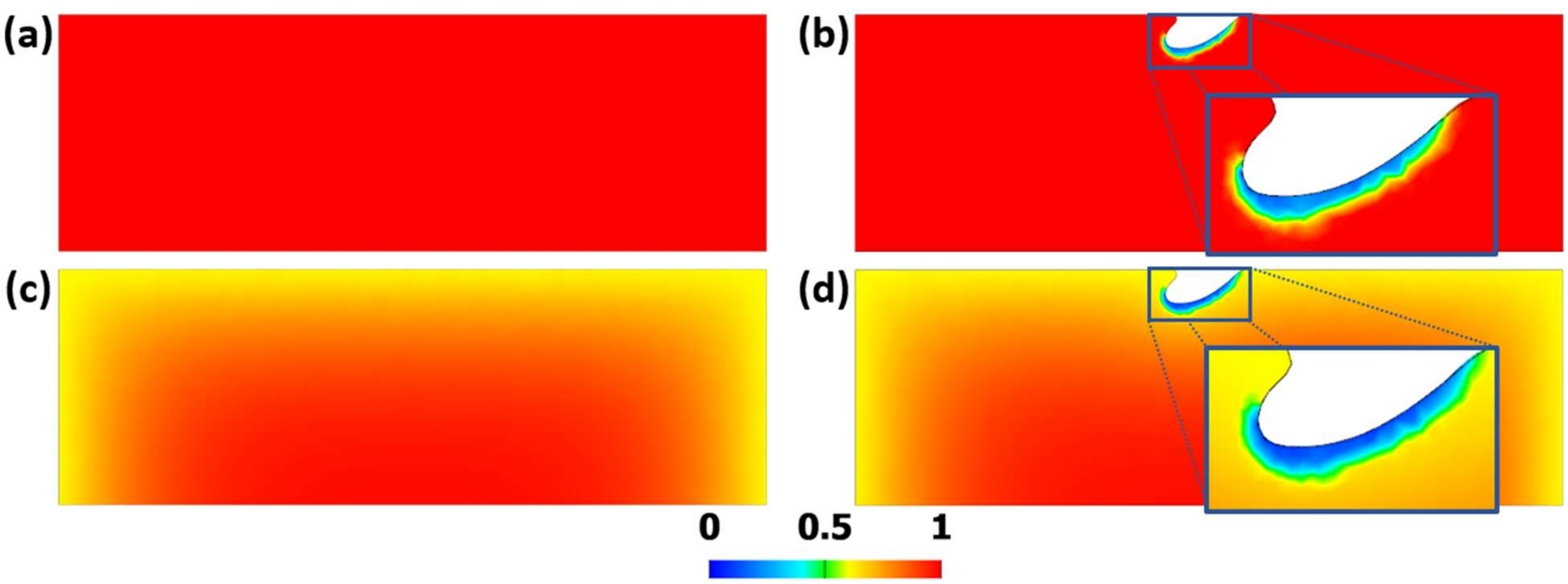

**Figure 4**: Normalized proteoglycan concentration at day 4 post-injury. (a) The mechanical-only model of the intact cartilage shows no PG degradation. (b) The injury model shows some PG loss under the lesion due to high shear strain. (c) When the intact model was subjected to IL-1 diffusion, minimum PG loss was observed near the outer surfaces. (d) With combination of mechanical trigger and pro-inflammatory cytokine influx, larger PG loss was seen near the lesion after 4$^{th}$ day of injury.

### 3.4 Example Problem: Bone Growth and Remodeling

This example demonstrates the use of FUSE to couple a reaction-diffusion model of cell-mediated bone formation with a short-time-scale mechanical model of a healing long-bone defect. The example was motivated by computational and experimental studies showing that fracture healing is regulated by the combined effects of progenitor cell migration, extracellular matrix production and mechanical stimulation [41-44]. This problem included three reacting species, multiple field exchanges, state-dependent material-property updates, and a time-dependent loading protocol. The objective was to demonstrate how FUSE can coordinate mechanobiological feedback between independently defined chemical and mechanical FEBio models during an 84-day fracture-healing simulation.

The mechanical model was used as the secondary model. The geometry represented a quarter-symmetric rat femur defect callus adapted from the fractured femur geometry used in previous studies, but with a more anatomically detailed external callus profile (Fig. 5a) [43, 44]. The computational domain included a central defect/callus, marrow and periosteal surfaces, and the surrounding muscle boundary. The callus initially consisted of soft granulation tissue with an elastic modulus of 0.05 MPa and a Poisson's ratio of 0.1 [43]. Newly formed bone was assigned a Poisson's ratio of 0.3, and its elastic modulus was updated from the local bone-matrix concentration using a mapping (Fig. 5b). This mapping used a pointwise relationship between normalized bone matrix concentration and elastic modulus [45]. The bottom face of the quarter-symmetry defect model was fixed in all directions (Fig. 5a). A pressure load was applied at the intact bone-callus contact surface. The loading protocol was motivated by an experimental study showing that cyclic axial compression applied after four days of healing enhanced fracture healing at relatively low load magnitude [46]. In the present model, the applied pressure was therefore kept negligible during the first four days after injury, increased progressively during early healing, reaching 50 MPa by day 35, and subsequently was held at that value until day 84 (Fig. 5c). In the FUSE control file, this was implemented as a *parameter-to-parameter* exchange from chemical-model time to mechanical-model pressure (Fig. 5c). At each daily coupling step, the mechanical problem was solved for two hours to estimate the mechanical state associated with the applied loading of that day. The primary mechanobiological feedback was implemented through the 3rd principal Lagrange (compressive) strain. The mechanical model transferred the compressive strain to the chemical model as the field $k_{osteo}$, which controlled the rate of MSC differentiation into osteoblasts (Eqn. 16). The implemented filter allowed MSC differentiation when compressive strain was between 0.5-5%, and zero outside this osteogenic window (Fig. 5d) [47]. A strain-dependent effective rate also represented bone resorption, $k_{resorp}$ (Eqn. 20). This exchange in the FUSE control file mapped compressive strain to $k_{resorp}$, with a maximum resorption rate of 0.07

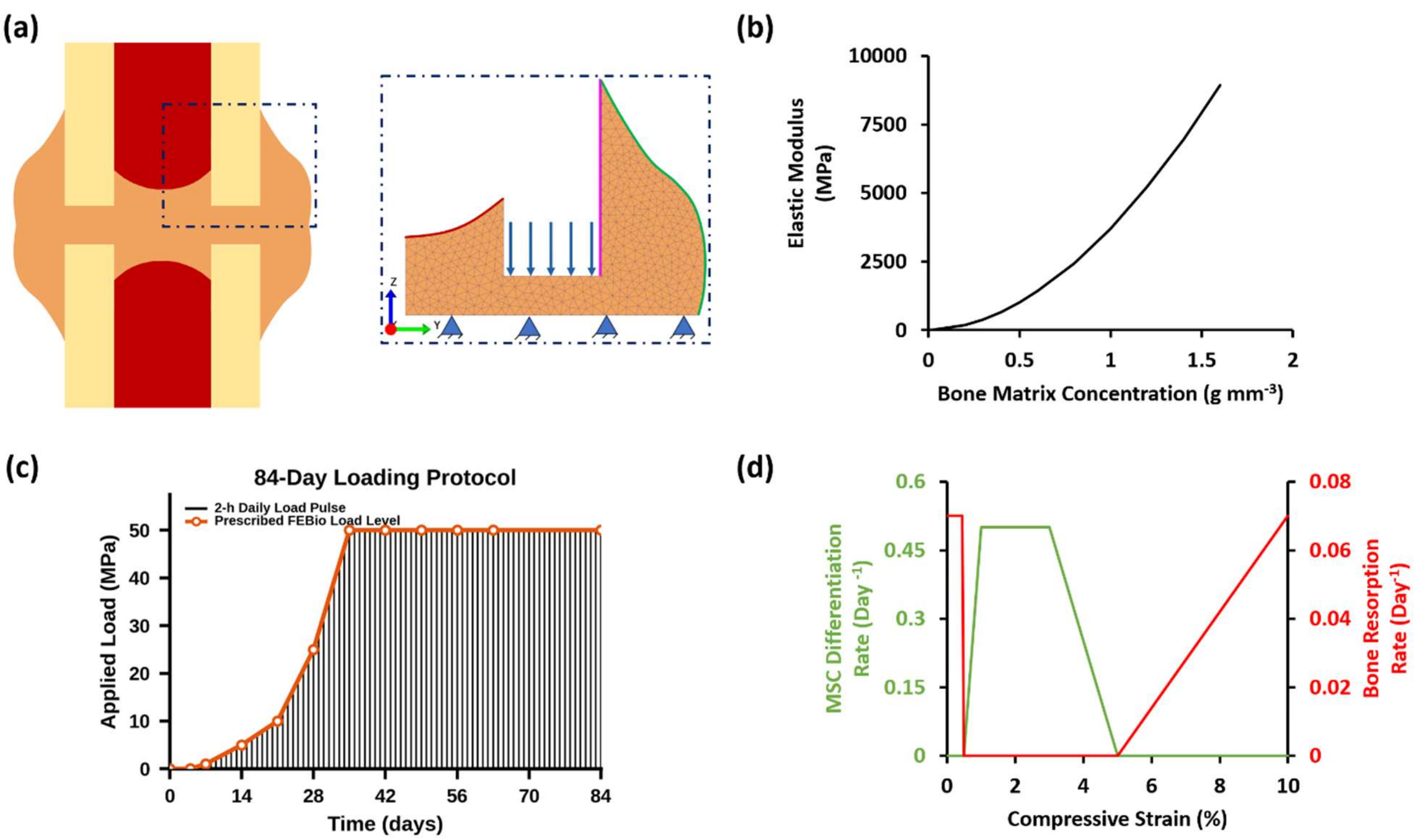

**Figure 5**: Coupled FUSE workflow for simulating mechanically regulated bone growth and remodeling after fracture. (a) Schematic of the rat femur defect model with bone (yellow), marrow (red), and callus (orange). The zoomed quarter-symmetric domain shows the meshed callus geometry, fixed symmetric surface, and pressure load applied at the intact bone-callus contact surface. The red outline denotes the bone marrow region, the magenta line indicates the periosteal surface, and the green line represents the muscle-facing surface. (b) Mapping between local bone matrix concentration and elastic modulus used to update the mechanical material properties as new bone matrix accumulated. (c) 84-day loading protocol implemented using a parameter-to-parameter FUSE exchange, where the prescribed FEBio load increased during early healing and was applied as a 2-h daily mechanical load pulse (dark lines) followed by chemical reaction/healing for the remainder of the day (white space). (d) Strain-dependent mechanobiological filters used to regulate MSC differentiation and bone resorption. MSC differentiation was activated within the osteogenic compressive-strain window, while bone resorption was prescribed at strains outside the osteogenic range.

day$^{-1}$ when compressive strain was below 0.5% or above 5%, and no resorption in the intermediate strain range (Fig. 5d).

The chemical model was used as the primary model and solved over 84 days with a one-day exchange interval. MSCs were modeled as a diffusing species with a diffusivity of 0.1715 mm²/day during the first seven days after fracture [43]. The MSC concentration was normalized such that the periosteal and marrow boundaries supplied MSCs at a concentration of 1, while the muscle-facing boundary supplied MSCs at a reduced concentration of 0.25. Bone marrow and periosteum were prescribed as the dominant MSC source boundaries, while the muscle-facing boundary was assigned a lower MSC supply. This assumption is consistent with lineage-tracing studies showing that bone marrow and periosteal skeletal stem/progenitor cells make major contributions to bone maintenance and repair [48]. Additional studies support the periosteum as a highly regenerative skeletal progenitor niche and a mechanosensitive source of osteoblast-lineage cells during bone formation [49, 50]. Muscle-resident progenitors can also contribute to fracture repair, particularly to the outer callus adjacent to muscle, but their contribution is more spatially restricted and injury-context dependent than the periosteal and marrow sources [51]. Diffusing MSCs could differentiate into osteoblasts, which were treated as solid-bound molecules. Osteoblasts then secreted bone matrix, also modeled as a solid-bound molecule. The chemical model, therefore, represented a simplified intramembranous bone-formation pathway consisting of MSC diffusion, strain-regulated osteoblast differentiation, MSC loss, osteoblast loss, bone-matrix secretion, and bone-matrix resorption (Eqns. 16-20):

$$MSC \xrightarrow{k_{osteo}} Osteoblast\,, \tag{16}$$

$$MSC \xrightarrow{k_{MSC}} [\ ], \tag{17}$$

$$Osteoblast \xrightarrow{k_{apoptosis}} [\ ], \tag{18}$$

$$Osteoblast \xrightarrow{k_{bone}} Bone\ Matrix + Osteoblast\ , \tag{19}$$

$$Bone\ Matrix \xrightarrow{k_{resorp}} [\ ]. \tag{20}$$

Here, $k_{osteo}$ and $k_{resorp}$ were mapped from mechanical stimulus, as discussed above. Osteoblast apoptosis was assigned a rate of $k_{apoptosis}$ = 0.075 day$^{-1}$, based on values used in previous studies [44]. Bone matrix secretion rate, $k_{bone}$ = 0.07 day$^{-1}$, was selected by parameter iteration to obtain a reasonable temporal progression of bone formation while keeping it similar to the mechanical bone-synthesis rate used in previous studies (0.05 day$^{-1}$) [43]. Finally, the MSC death was assigned a first-order rate of $k_{MSC}$ = 0.1 day$^{-1}$. This term was included as a simple sink because the model did not include MSC proliferation, differentiation into chondrocytes or fibroblasts, or recruitment regulated by growth factors [41, 42, 44].

The coupled simulation produced progressive bone-matrix accumulation and a corresponding increase in elastic modulus over the 84-day healing period. Bone matrix was low throughout the callus at day 1 (Fig. ***6***). By day 14, new matrix began to appear within the defect, indicating the onset of osteoblast-mediated bone formation. Between days 21 and 56, bone matrix increased substantially in the central defect and near the regions where the intact bone contacted the callus (Fig. ***6***). At later time points, matrix that formed along the outer skirt of the callus was partially resorbed, while the central defect region retained high bone-matrix concentration. Because mechanical stiffness was updated directly from bone matrix, the elastic modulus field followed the same spatial pattern (Fig. ***6***). Stiffness increased first in regions of early bone formation and became highest by days 56-84 in the regions with sustained matrix accumulation. The resulting spatial evolution demonstrates two-way mechanobiological feedback: strain fields regulate MSC differentiation into osteoblasts, osteoblasts produce bone matrix, bone matrix increases local stiffness, and the updated stiffness field alters the subsequent strain environment.

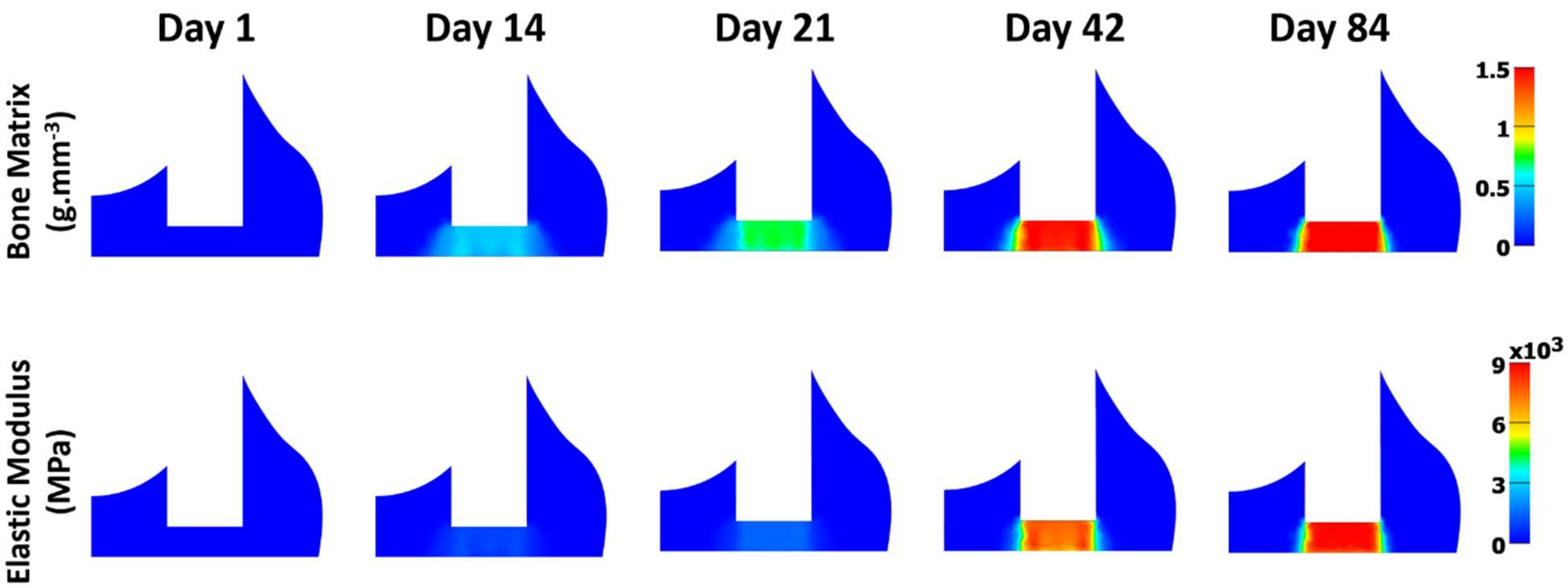

**Figure 6**: Temporal evolution of bone matrix deposition and stiffness recovery during the coupled fracture-healing simulation. The top row shows bone matrix concentration in the quarter-symmetric callus domain at days 1, 14, 21, 42, and 84. Bone matrix was initially low throughout the defect, began to appear within the callus by day 14, and progressively accumulated in the central defect region and near the intact bone-callus interfaces. The bottom row shows the corresponding elastic modulus field, which was updated from the local bone matrix concentration at each coupling step. As bone matrix accumulated, the callus stiffness increased spatially, with the highest elastic modulus developing by day 84 in regions of sustained bone formation.

This example demonstrates how FUSE can coordinate biological adaptation and mechanical analysis within a complex mechanobiology workflow. In this setup, a long-term biological model drives the overall simulation progression, while a short-duration mechanical model is repeatedly called to update the evolving mechanical environment. Through this coupling, the workflow transfers biologically meaningful quantities between models, including compressive strain, reaction-rate maps, bone-matrix concentration, elastic modulus, and applied load. These exchanges include both spatial field mappings and global parameter updates, enabling local tissue properties and overall loading conditions to evolve together over time. The example also illustrates how point filters can be used to encode threshold-based mechanobiological rules, such as osteogenic and resorptive strain windows, without requiring changes to the underlying FEBio solvers. Taken together, this example serves as a demonstration of FUSE for coordinating multiphysics, multiscale model interactions in complex mechanobiological processes, rather than as a validated predictive model of rat femur repair.

## 4. Discussion

This work presents FUSE as a partitioned coupling plugin for coordinating separately defined FEBio physics models through reusable field exchange. FUSE provides an execution framework that allows existing FEBio models to exchange spatially varying quantities, apply user-defined filters, and advance in a coordinated workflow. This addresses a common need in computational biomechanics, where mechanical, chemical, and biological subsystems are often formulated separately but interact through feedback across different temporal scales [10, 12]. By building on the existing plugin architecture, data maps, and multiphysics infrastructure in FEBio, FUSE extends the ecosystem without requiring users to reformulate coupled problems monolithically [13, 15, 19].

The two verification studies illustrate that the partitioned implementation reproduces known solutions with high accuracy and demonstrate that the core exchange operations are implemented correctly. Together, these verification cases support the numerical correctness of the FUSE workflow before applying it to a more complex mechanobiological example. The cartilage degradation and fracture healing examples illustrate the intended use of FUSE rather than serving as biological validation. The cartilage example combines a short-time-scale mechanical analysis with a longer-time-scale reaction-diffusion model using repeated field exchange, while the fracture-healing example extends this workflow by incorporating multiple reacting species, strain-regulated cell differentiation and resorption, evolving material properties, and a time-dependent loading protocol. Although both examples are simplified relative to published biological models, they demonstrate how independently developed FEBio models can be assembled into coupled mechanobiological simulations. These examples highlight the types of workflows that motivated the development of FUSE, including applications involving tissue degradation, growth and remodeling, tissue engineering, mechanochemical signaling, and multiscale biology.

The current implementation has several limitations that suggest directions for future development. First, FUSE presently assumes identical meshes and matching node and element ordering between coupled models. This simplifies direct data transfer but limits applications requiring different discretizations for different physics. We anticipate relaxing this restriction in the near future. Second, by design, the current workflow is time-decoupled, with the secondary models solved at prescribed exchange intervals rather than advancing synchronously with the primary model. This is useful for problems with separated time scales, but strongly coupled applications may require synchronized time advancement or iterative coupling within each global step. Third, the current implementation transfers the final output state of each secondary model. Future versions could support temporal reductions such as peak, average, minimum, maximum, or accumulated quantities. Finally, when there are multiple secondary models, they are currently solved sequentially. Independent secondary analyses could easily be parallelized in future implementations.

Overall, FUSE provides a practical software infrastructure for partitioned multiphysics simulation within the FEBio ecosystem. By separating model coordination from the underlying physics implementations, it enables independently developed mechanics, transport, chemistry, and biological models to participate in reproducible coupled simulations while remaining individually maintainable and extensible. We anticipate that this capability will facilitate the development of increasingly sophisticated multiscale biomechanical models while allowing researchers to leverage the growing collection of existing FEBio physics modules and plugins.